\documentclass{PoS}
\usepackage{times}
\usepackage{amsfonts}
\usepackage{amssymb}
\usepackage{amsmath}
\usepackage{stmaryrd}
\usepackage{amscd}
\usepackage{latexsym}
\usepackage{bbm}
\def\a{\alpha}
\def\b{\beta}
\def\g{\gamma}

\def\de{\delta}
\def\eps{\epsilon}
\def\la{\lambda}
\def\m{\mu}
\def\n{\nu}
\def\th{\theta}
\def\sfrac#1#2{{\textstyle\frac{#1}{#2}}}

\def\ph{\phantom{-}}

\def\diff{\mathrm{d}}
\def\tr{\mathrm{tr}}
\def\pa{\partial}
\def\>{\rangle}
\def\<{\langle}
\def\+{\dagger}
\def\={\ =\ }
\def\Tdag{T^{\dagger}}
\def\zb{{\bar z}}

\newcommand{\C}{\mathbb C}
\newcommand{\R}{\mathbb R}

\newcommand{\unity}{\mathbbm{1}}
\newcommand{\be}{\begin{equation}}
\newcommand{\ee}{\end{equation}}
\newcommand{\bea}{\begin{eqnarray}}
\newcommand{\eea}{\end{eqnarray}}
\newcommand{\bal}{\begin{aligned}}
\newcommand{\eal}{\end{aligned}}
\newcommand{\non}{\nonumber}
\newcommand{\und}{\qquad\textrm{and}\qquad}

\title{Noncommutative deformation of the Ward metric}

\ShortTitle{Noncommutative Ward metric}

\author{Magnus Goffeng\\
Institut f\"ur Analysis,
Leibniz Universit\"at Hannover, 
30167 Hannover, Germany \\
E-mail: \email{ goffeng@math.uni-hannover.de}}

\author{\speaker{Olaf Lechtenfeld}
\thanks{The talk is based on \cite{lema} and subsequent joint work of the authors.}\\
Institut f\"ur Theoretische Physik,
Leibniz Universit\"at Hannover, 
30167 Hannover, Germany \\
E-mail: \email{ lechtenf@itp.uni-hannover.de}}

\abstract{
The moduli-space metric in the static non-Abelian charge-two sector 
of the Moyal-deformed $\C P^1$ sigma model in $1{+}2$ dimensions is analyzed. 
After recalling the commutative results of Ward and Ruback 
and the $\zeta$-regularized construction of the noncommutative K\"ahler potential 
due to the second author, explicit expressions and asymptotics for it
are presented and discussed in different regions of the moduli space. 
Along two curves in the moduli space the potential can be calculated analytically. 
In the region of solitons known as ``ring-like'', perturbation theory is used. 
In the region of \ ``lump-like'' solitons, both perturbation theory 
and the $\zeta$-function approach are employed. While the strong noncommutativity
limit is smooth and under control, the commutative limit in the two-lump region
remains a semiclassical challenge.
}

\FullConference{
Proceedings of the Corfu Summer Institute 2011 School and Workshops on Elementary Particle Physics and Gravity\\
September 4-18, 2011\\
Corfu, Greece}

\begin{document}

\section{Introduction}

\noindent
The metric structure of the moduli space of solutions to the $\C P^1$ model was first studied by Ward \cite{ward}. It was later shown by Ruback \cite{ruback} that this metric comes from a K\"ahler potential. Formal integration of the energy functional gives the potential as a certain integral. By Moyal deformation and replacing the integral by a $\zeta$-regularized trace, the second author \cite{lema} introduced a noncommutative deformation of the K\"ahler potential. The moduli space at topological charge two contains two interesting regions called the ring regime and the two-lump regime, motivated by the form of the energy densities for the corresponding solitons. In the ring regime, the behavior of the deformed K\"ahler potential is known from \cite{lema}. 

In this paper we review the results of~\cite{lema} and explore the metric structure in the two-lump regime further. We apply two different techniques, perturbation theory and an explicit calculation involving a $\zeta$-function. The first approach relies on the solution of a singular Sturm-Liouville problem and gives the asymptotics of the K\"ahler potential in the strong noncommutative limit. In the second approach we calculate the $\zeta$-function of the involved operator explicitly. We compare the two approaches at strong noncommutativity.

The paper is organized as follows. The $\C P^1$ model and its moduli-space metric are briefly reviewed in Sections 2 and~3, respectively. In Section~4 we recall some standard facts about Moyal deformation and $\zeta$-regularization. In Section 5 we present the main results from~\cite{lema} on the deformed potential in the ring regime. Finally, in Sections 6 and~7 we calculate the asymptotics in the noncommutative limit for the two-lump regime using perturbation theory respectively $\zeta$-functions.
 
\section{The $\C P^1$ model and its solitons}

\noindent
The $\C P^1$ sigma model in $1{+}2$ dimensions is a paradigm for soliton
studies~\cite{zakr,mantonbook}. It describes the dynamics of maps 
\be
u: \ (t,z,\zb)\in\R^{1,2}\quad\longrightarrow\quad
S^2\simeq\sfrac{\textrm{SU}(2)}{\textrm{U}(1)}\simeq \C P^1 \ .
\ee
It is useful to introduce homogeneous complex coordinates via $u=\sfrac{p}{q}$, 
so that~\footnote{
We restrict ourselves to one of two patches covering $S^2$. This is inessential here.}
\be
T\=\bigl(\begin{smallmatrix} p \\[4pt] q \end{smallmatrix}\bigr)
\ \sim\ \bigl(\begin{smallmatrix} u \\[4pt] 1 \end{smallmatrix}\bigr)
\qquad\Longrightarrow\qquad
P\=P^\+\=P^2\=T\,\sfrac{1}{\Tdag T}\,\Tdag\ ,
\ee
and hermitian rank-one projectors in $\C^2$ appear.

The model is defined by its action functional,
\be
\bal
S&\=-4\int\!\diff^3x\;\tr\,\eta^{\m\n}\pa_\m P\,\pa_\n P \\[4pt]
&\=-4\int\!\diff^3x\;(\Tdag T)^{-1}\eta^{\m\n}\pa_\m\Tdag(\unity{-}P)\,\pa_\n T
\qquad\qquad\qquad\qquad\qquad\qquad\ph\\[4pt]
&\=-4\int\!\diff^3x\;(1{+}\bar uu)^{-2}\eta^{\m\n}\pa_\m\bar u\,\pa_\n u\ ,
\eal
\ee
which for static configurations, $\pa_t u=0$, reduces (up to a range-of-$t$ factor) 
to the energy functional
\be
\bal
E&\=8\int\!\diff^2z\;(\Tdag T)^{-1}
\bigl\{\pa_\zb\Tdag(\unity{-}P)\,\pa_zT\ +\ \pa_z\Tdag(\unity{-}P)\,\pa_\zb T\bigr\} \\[4pt]
&\=8\int\!\diff^2z\;(1{+}\bar uu)^{-2}
\bigr\{\pa_\zb\bar u\,\pa_zu+\pa_z\bar u\,\pa_\zb u\bigr\}\ .
\eal
\ee
Classical static configurations are those which minimize~$E$. 
Obviously, any meromorphic $u=u(z)$ or anti-meromorphic $u=u(\zb)$ is classical.
Furthermore, $u$ must be a rational function (of some degree $n$) for the energy to be finite.
In this case, one has $E=8\pi|n|$, and one speaks of solitons (or anti-solitons).
The moduli space of such soliton solutions has complex dimension~$2n{+}1$.
After quotienting the moduli space by the action associated with domain and target isometries, 
a reduced moduli space~${\cal M}_n$ parametrized by
a set $\{\a\}=\{\b,\g,\de,\eps,\ldots\}$ of nontrivial moduli remains. For example:
\be
n{=}1:\quad
T(z)\=\Bigl(\begin{matrix} \b \\ z \end{matrix}\Bigr)
\qquad\qquad\Rightarrow\quad
E\=\int\!\diff^2z\;\frac{8\,\b^2}{(\b^2+|z|^2)^2}
\=8\pi\ ,\qquad\qquad\!\!\ph
\ee
\be
n{=}2:\quad
T(z)\=\Bigl(\begin{matrix} \b z+\g \\ z^2+\eps \end{matrix}\Bigr)
\qquad\Rightarrow\quad
E\=\int\!\diff^2z\;\frac{8\,|\b z^2+2\g\,z-\b\eps|^2}{(|\b z+\g|^2+|z^2+\eps|^2)^2}
\=16\pi\ ,
\ee
with $\b,\g\in \R_{\ge0}$ and $\eps\in\C$,
so that $\textrm{dim}_\R{\cal M}_1=1$ and $\textrm{dim}_\R{\cal M}_2=4$.
In this paper, we choose to specialize to the subclass $\b{=}0$,
which allows one to also rotate away the phase of~$\eps$:
\be
T(z)\=\Bigl(\begin{matrix} \g \\ z^2+\eps \end{matrix}\Bigr)
\quad\Rightarrow\quad
E\=\int\!\diff^2z\;\frac{32\,|\g\,z|^2}{(\g^2+|z^2+\eps|^2)^2}
\=16\pi \qquad\textrm{with}\quad \g\in\R_{>0},\ \eps\in\R_{\ge0}\ .
\ee
Note that the value $\b=0$ should be excluded of ${\cal M}_1$,
and $\g=0$ is not part of ${\cal M}_2^{\b=0}$.

\section{Moduli space metric}

\noindent
The $\C P^1$ model provides the simplest example for a nontrivial dynamics 
of moving lumps. For sufficiently slow motion, we can apply the
adiabatic approximation scheme of Manton~\cite{manton}.
So far we considered classical static finite-energy solutions (solitons)
$u=u(z\,|\,\a)$ depending on moduli parameters~$\a$.
The adiabatic or moduli-space approximation brings back the time
dependence as a sequence of snapshots of static solitons,
\be
u(t,z,\zb)\ \approx\ u(z\,|\,\a(t)) \ ,
\ee
which pushes the true trajectory into the static soliton moduli space.
We may regard all moduli as complex numbers.
Evaluating the action on this moduli-space trajectory yields
\be
\bal
S\bigl[u(\cdot\,|\,\a(t))\bigr] \ +\ \textrm{const}
&\=4\int\!\diff{t}\;\Bigl[ \smallint\diff^2z\;
(\Tdag T)^{-1}\pa_{\bar\a}\Tdag(\unity{-}P)\,\pa_\a T \Bigr]\,
|\dot\a|^2 \\[4pt]
&\=4\int\!\diff{t}\;\Bigl[ \smallint\diff^2z\;
\sfrac{\pa_{\bar\a}\bar u\,\pa_\a u}{(1{+}\bar uu)^2}\Bigr]\,
|\dot\a|^2
\ =:\ \sfrac12\int\!\diff{t}\;g_{\bar\a\a}(\a)\,|\dot\a|^2\ ,
\eal
\ee
which defines a K\"ahler metric \
$g_{\bar\a\a}=\pa_{\bar\a}\pa_\a{\cal K}$
on the moduli space. Extremizing this moduli-space action yields
geodesic motion for this metric. The corresponding K\"ahler potential reads
\be \label{Kdefcom}
{\cal K}\=8\int\!\diff^2z\;\ln(1+\bar uu)
\=8\int\!\diff^2z\;\ln(1+\sfrac{\bar pp}{\bar qq})
\=8\int\!\diff^2z\;\bigl[\ln \Tdag T - \ln \bar qq\bigr]\ .
\ee
Some moduli have infinite inertia 
and must be treated as external parameters. At $n{=}1$ and $n{=}2$ 
this is the case for $\b$, for which one finds that
$g_{\bar\b \b }=\pa_{\bar\b}\pa_\b{\cal K}=\infty$.

In the $n{=}1$ sector, the K\"ahler potential is divergent,
but can be regularized by differentiating twice under the integral
with respect to a regularization parameter~$\de$,
\be \label{n1com}
T=\bigl(\begin{smallmatrix} \b \\ z+\de \end{smallmatrix}\bigr)
\quad\Rightarrow\quad
\pa_{\bar\de}\pa_\de{\cal K}\ :=\ 
8\int\!\diff^2z\;\pa_{\bar\de}\pa_\de\ln\bigl(1+\sfrac{|\b|^2}{|z+\de|^2}\bigr)
\=8\pi 
\quad\Rightarrow\quad
{\cal K}\ \buildrel{\textrm{reg}}\over{=}\ 8\pi\bar\de\de\ .
\ee
This is the expected form for the uniform motion of a single lump
in the $\de$~plane.

In contrast, for $n{=}2$ the K\"ahler potential is well defined but nontrivial,
\be
T(z)\=\bigl(\begin{smallmatrix} \g \\ z^2+\eps \end{smallmatrix}\bigr)
\quad\Rightarrow\quad
{\cal K}\=8\int\!\diff^2z\;\ln\bigl(1+\sfrac{|\g|^2}{|z^2+\eps|^2}\bigr)\=
16\pi\,|\g|\,E\bigl(-|\sfrac\eps\g|^2\bigr)\ =:\ {\cal K}_0(\g,\eps)\ ,
\ee
where $E(m{=}k^2)$ denotes the complete elliptic integral of the second kind.
This result was first obtained by Ruback~\cite{ruback} (see also~\cite{dunajski}),
after the metric and the geodesic motion had been analyzed earlier by Ward~\cite{ward}.

Depending on the relative size of the two dimensionful moduli, $|\g|$ and $|\eps|$,
we can distinguish two (asymptotic) types of solitons in~${\cal M}_2^{\b=0}$:
For $|\eps|\ll|\g|$ the energy density is localized in a ring-like region in the $z$-plane,
while for $|\g|\ll|\eps|$ it sharply peaks at the two locations $z_\pm=\pm\sqrt{-\eps}$.
In between, the configurations are of intermediate type. In this `ring' and `two-lump'
regimes of~${\cal M}_2^{\b=0}$, the K\"ahler potential admits an expansion in the small ratio,
{\small
\bea
\frac{{\cal K}_0}{16\pi}&=&\frac{\pi}{2}\,|\g|\,\Bigl\{ 1+ \frac14\Bigl|\frac\eps\g\Bigr|^2
-\frac3{64}\Bigl|\frac\eps\g\Bigr|^4 +\frac5{256}\Bigl|\frac\eps\g\Bigr|^6
+\ldots \Bigr\} \label{Kringcom}
\qquad\textrm{\normalsize and} \\ 
\frac{{\cal K}_0}{16\pi}&=&|\eps|\,\Bigl\{ 1-
\frac14\Bigl(-1+\ln\bigl|\frac\g{4\eps}\bigr|^2\Bigr)
\Bigl|\frac\g\eps\Bigr|^2+
\frac1{32}\Bigl(\frac32+\ln\bigl|\frac\g{4\eps}\bigr|^2\Bigr)
\Bigl|\frac\g\eps\Bigr|^4-
\frac3{256}\Bigl(2+\ln\bigl|\frac\g{4\eps}\bigr|^2\Bigr)
\Bigl|\frac\g\eps\Bigr|^6+\ldots\Bigr\}\ ,\quad\ph \label{Klumpcom}
\eea
}
respectively. For $|\g|\to0$ (at the boundary of ${\cal M}_2^{\b=0}$)
one encounters mild logarithmic singularities.
For illustration, we display some energy density plots for the adiabatic motion 
in the ring regime.

\begin{figure}
\includegraphics[width=7cm]{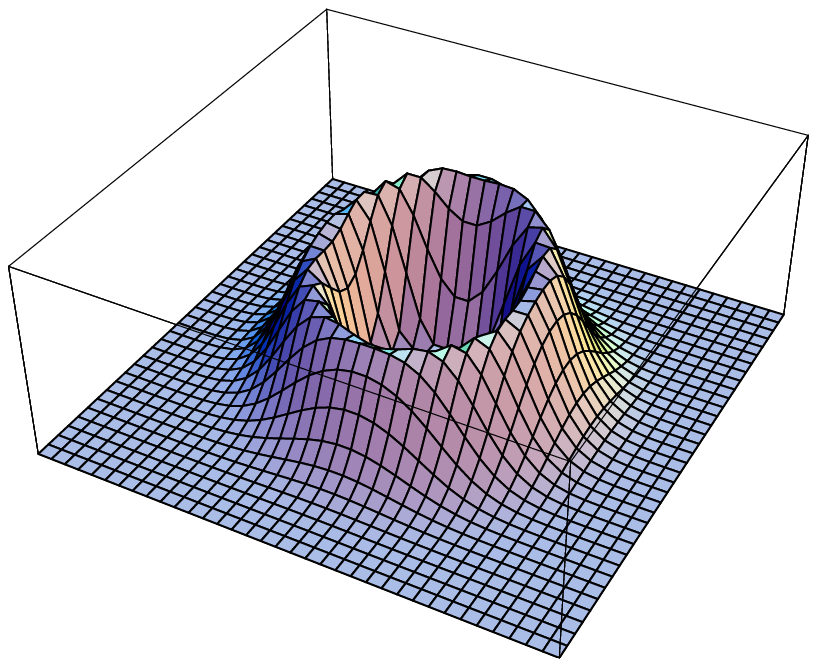} \qquad
\includegraphics[width=7cm]{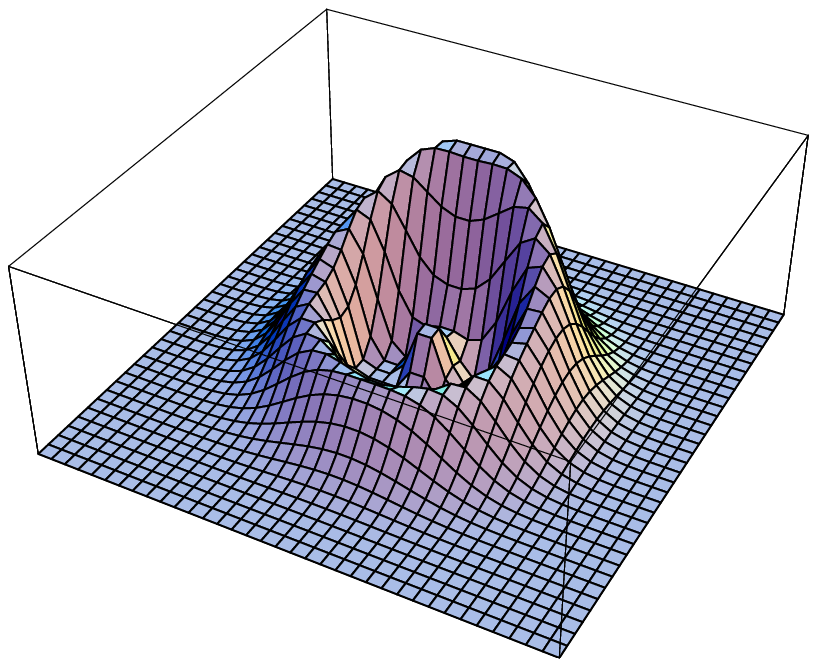} \\
\includegraphics[width=7cm]{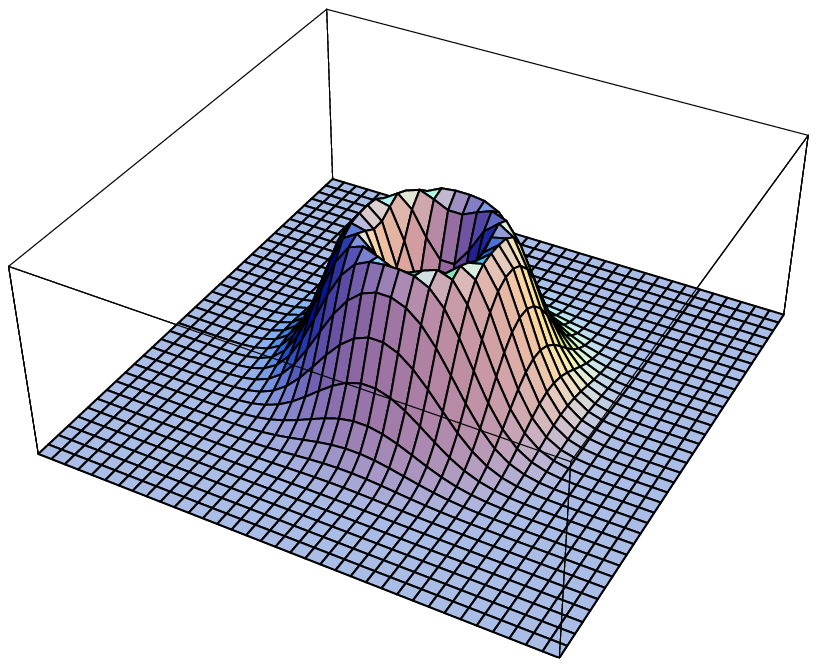} \qquad
\includegraphics[width=7cm]{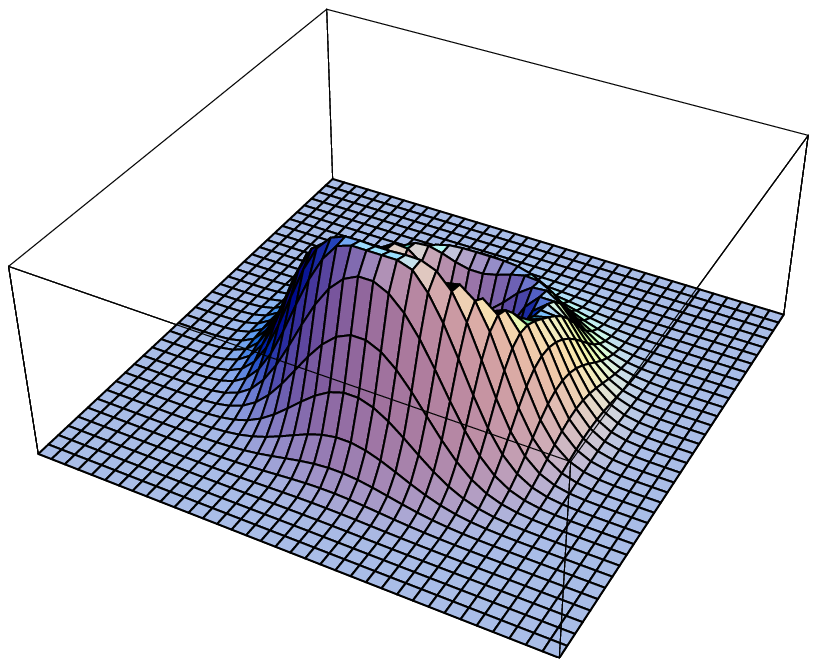}
\caption{Adiabatic motion in ring regime: energy density plots
for $t$-values $-5.0$, $-1.5$, $0.0$ and $1.5$}
\label{fig1}
\end{figure}

\section{Moyal deformation}

\noindent
We proceed to the noncommutative generalization of the $\C P^1$ model~\cite{lee,furuta}.
Loosely speaking, the Moyal deformation replaces the spatial coordinates $(z,\zb)$ by
operators $(Z,Z^\+)$ subject to the Heisenberg-algebra commutation relation
$[Z,Z^\+]=2\th=\textrm{const}$. A standard physics realization in terms of semi-infinite
matrices reads
\be
Z\=\sqrt{2\th}\,a\=\sqrt{2\th}{\left( \begin{smallmatrix}
0 & 0 &&& \\[2pt] \sqrt{1} & 0 & 0 && \\[0pt] & \sqrt{2} & 0 & 0 & \\[-4pt]
&& \sqrt{3} & 0 & \ddots \\[-4pt] &&& \ddots & \ddots \end{smallmatrix} \right)}
\quad\&\quad
Z^\+\=\sqrt{2\th}\,a^\+\=\sqrt{2\th}{\left( \begin{smallmatrix}
0\,& \sqrt{1}&&& \\[2pt] 0 & 0 & \sqrt{2}&& \\[0pt] & 0 & 0 & \sqrt{3}&\\[-4pt]
&& 0 & 0 & \ddots \\[-4pt] &&& \ddots & \ddots \end{smallmatrix} \right)}\ .
\ee
This quantization map is extended linearly and associatively to a large class of
functions~$f$ of $(z,\zb)\in\C$ via
\be
f \quad\longmapsto\quad F(f)\=f(Z,Z^\+)\big|_{\textrm{sym}}\ ,
\ee
where the symmetric (or Weyl) operator ordering of monomials is indicated.
In the $\C P^1$ model, examples for $f$ are the polynomials $p$ and $q$, 
so $f$ and $F$ take value in $\C P^1$.
Partial derivatives also deform easily,
\be
\pa_\zb\quad\mapsto\quad
\sfrac1{2\th}[Z,\,\cdot\,]\=\sfrac1{\sqrt{2\th}}[a,\,\cdot\,] \und
\pa_z \quad\mapsto\quad
-\sfrac1{2\th}[Z^\+,\,\cdot\,]\=-\sfrac1{\sqrt{2\th}}[a^\+,\,\cdot\,]\ .
\ee
The Heisenberg algebra is represented (with highest weight $|0\>$) on the Fock space~$\cal F$,
\be
[\,a,a^\+]=\unity \quad\textrm{and}\quad
a\,|0\>=0 \qquad\Rightarrow\qquad {\cal F}\=\text{span}
\bigl\{|n\>=\sfrac1{\sqrt{n!}}(a^\+)^n|0\>\ \bigm|\ n=0,1,2,\ldots\bigr\}\ ,
\ee
where we have introduced an eigenbasis of the number operator
\be
N=a^\+a \qquad\Rightarrow\qquad
N\,|n\>\=n\,|n\> \und \<n|n\>\=1 \qquad\textrm{for}\qquad n=0,1,2,\ldots\ .
\ee

Let us remark that the operator $a$ can also be realized as the unbounded operator $a=\frac{1}{\sqrt{2}}(\partial_x+x)$ on $L^2(\R)$. This operator is an operator of order $1$ in the Shubin calculus, see Chapter IV of \cite{shubin}, so any noncommutative polynomial in $a$ and $a^\dagger$ is an operator in the Shubin calculus. The Shubin calculus allows one to apply pseudo-differential techniques to the involved operators guaranteeing good spectral properties of elliptic operators and also asymptotics of parameters with semi-classical methods, in principle an elliptic operator in the Shubin calculus behaves as an elliptic operator on a compact manifold of twice the dimension. In particular, this calculus provides the mathematical machinery for the quantization map associating an operator $F(f)$ with a function $f$ on~$\C$. Under suitable assumptions on $f$, such as Schwartz class, the integral over~$\C$ deforms to a trace over the Fock space~$\cal F$, and one has that
\be
2\pi\theta\,\tr\, F(f)\= \int\!\diff^2 z\ f(z,\bar{z}) \und
2\pi\theta\,\tr\, f(Z,Z^\+) \= \int\!\diff^2 z\ f_\star(z,\bar{z})\ ,
\ee
where the star indicates the forming of monomials by using the Moyal star product.

We would like to define a Moyal-deformed K\"ahler potential~$\cal K$ as a potential for the noncommutative Ward metric.
This requires fixing the ordering ambiguity by making a particular choice.\footnote{
A symmetric ordering prescription corresponds to a zero deformation.}
A literal adaptation of~(\ref{Kdefcom}) fails, however, since $Q^\+Q=\bar{q}(Z^\+)q(Z)$ is in general not invertible,
and thus $U^\+U=\bar{u}(Z^\+)u(Z)$ does not exist. This difficulty is avoided by deforming instead
\be
8\int\!\diff^2z\ \ln (T^\+ T) \quad\longmapsto\quad
16\pi\,\th\,\tr\ln (T^\+ T) \= 16\pi\,\th\,\tr\ln (P^\+P+Q^\+Q)\ =:\ {\cal K}\ ,
\ee
where the entries of the deformed~$T:\C\times \mathcal{M}_n\to \C^2$ are of course operator-valued. 
However, like in the commutative case, this definition is only formal due to the lack of convergence.
The problem is that, since $T$ will be a differential operator, $\ln(T^\dagger T)$ is not of trace class, 
and so the expression does not make sense. 

To deal with this type of problems the technique of $\zeta$-regularization was invented by Ray and Singer~\cite{raysinger}. The idea behind $\zeta$-regularization is that $\ln(x)=-\frac{\diff}{\diff s}\big|_{s=0} x^{-s}$. So formally $\sum_{x}\ln(x)=-\frac{\diff}{\diff s}\big|_{s=0} \sum_x x^{-s}$, and one can hopefully make sense of $\sum_x x^{-s}$ as a holomorphic function at $s=0$ by a holomorphic extension. If $D$ is a positive operator, one can define $D^{-s}$ which, for $D$ being an elliptic differential operator, often is trace class at $\Re(s)$ large enough. If this is the case, one defines the $\zeta$-function of $D$ as 
\begin{equation}
\label{zetad}
\zeta_D(s)\ :=\ \tr (D^{-s})\ \equiv\ \frac{1}{\Gamma(s)}\int_0^\infty\!\diff t\ t^{s-1}\tr(\mathrm{e}^{-tD})\ ,
\end{equation}
which is a well defined holomorphic function for large $\Re(s)$. One can often extend this holomorphic function to a neighborhood of $0$ and define
\be
\tr\ln_\zeta(D)\ :=\ -\zeta_D'(0)\ .
\ee
Especially, if $D$ is an elliptic operator of order $m$ in the Shubin calculus, the expression \eqref{zetad} converges for $\Re(s)>2/m$. If it also depends on parameters (moduli), $\tr\ln_\zeta(D)$ admits asymptotic expansions both in the small and large parameter limits, see~\cite{burgheleafriedlanderkappeler}. The asymptotics at~$\infty$ only requires semiclassical information, but the asymptotics in~$0$ needs global information, i.e.~particular values of the $\zeta$-function.

Being interested in the parameter dependence of our operator we observe the functional equation
\begin{equation}
\label{functionalequation}
\frac{\diff}{\diff t} \zeta_{D+t}(s)\=-s\,\zeta_{D+t}(s{+}1) \qquad\textrm{for}\quad t>0\ .
\end{equation}
This regularizes the expression for the determinant since we can express the right hand side explicitly in a slightly larger domain. If $D$ is an order $m>2$ operator in the Shubin calculus, then
\be \label{functionalrelation}
\frac{\diff}{\diff t} \tr\ln_\zeta(D+t)\=\zeta_{D+t}(1)\=\tr\,(D+t)^{-1}\ .
\ee
Another useful result to calculate $\zeta$-determinants is a perturbation-type result valid for operators decomposing as $D=G+E$, where $G$ is invertible and $G^{-1}E$ is Schatten class. Then one has
\be \label{perturb}
\tr\ln(G+E)\=\tr\ln(G)\ +\ \tr\ln(\unity+G^{-1}E)\ .
\ee
The second term can sometimes be calculated in terms of an explicit series representation 
(see, e.g., Theorem~$4.3$ of~\cite{friedlandersthesis}). 

Let us specialize to the case at hand, $D=T^\+T=P^\+P+Q^\+Q$.
Anticipating (removable) zero-mode complications, we switch from solitons to antisolitons from now on,
i.e.\ take $T=T(Z^\+)$.
For $n{=}1$, one can proceed like in the commutative case and obtain a convergent trace after 
differentiating twice with respect to a shift parameter~$\delta$. Since the result equals $8\pi$
just as in~(\ref{n1com}), the K\"ahler potential is undeformed. This agrees with the expectation
for the dynamics of a single lump.

The $n{=}2$ sector provides the challenge we want to meet. 
Since the deformation parameter~$\th$ introduces a new scale into the problem, 
we can relate all dimensionful (greek) moduli to~$\th$ by introducing dimensionless (latin) moduli,
\be
Z=\sqrt{2\th}\,a\ ,\quad
\b=\sqrt{2\th}\,b\ ,\quad
\g=2\th\,g\ ,\quad
\eps=2\th\,e\ .
\ee
Note that, for fixed greek moduli, small latin moduli correspond to strong noncommutativity 
while the commutative limit is attained for infinitely large ones. In this notation, we have
\be
\bal
&T\=\Bigl(\begin{matrix} \bar g \\ a^{\+2}+\bar e \end{matrix}\Bigr)
\qquad\Rightarrow\qquad
{\cal K}\= 16\pi\th\,\tr\,\ln(\Tdag T) \qquad\qquad\text{with}\\[8pt]
&\Tdag T \= \bar gg\ +\ \underbrace{(a^2+e)(a^{\+2}+\bar e)}_F
\= \underbrace{\bar gg+(N{+}2)(N{+}1)}_G
\ +\ \underbrace{e\,a^{\+2} + \bar e\,a^2+\bar ee}_E \ ,
\eal 
\ee
introducing $F=T^\+T|_{g=0}$ and the decomposition into a $g$-dependent (diagonal)
and an $e$-dependent (non-diagonal) part of~$D$.
Important actions on the basis states are \
$G\,|n\>=\bigl[(n{+}2)(n{+}1)+\bar gg]|n\>$ \ and
\be
\nwarrow|n\>\ :=\ G^{-1}a^{\+2}|n\> \= \sfrac{\sqrt{(n{+}1)(n{+}2)}}{\bar gg+(n{+}4)(n{+}3)}\,|n{+}2\> 
\ ,\qquad
\swarrow|n\>\ :=\ G^{-1}a^2    |n\> \= \sfrac{\sqrt{n(n{-}1)}}{\bar gg+n(n{-}1)}\,|n{-}2\>\ .
\ee
The $\bar ee$ term may be omitted at first and produced at the end of the day by shifting 
$\bar gg\to\bar gg+\bar ee$.

\section{Deformed rings}

\noindent
For $|e|\ll|g|$ (the ring regime) it is reasonable to set up a perturbation expansion in
the off-diagonal part of the operator. We copy the formal method used in quantum field theory,
\be\label{ringpotential}
\bal
&\frac{\cal K}{16\pi\th}\=
\tr\,\ln G\ +\ \tr\,\ln(\unity+G^{-1} E)\= \tr\,\ln G \ -\
\sum_{k=1}^\infty\sfrac{(-1)^k}{k}\,\tr\,\bigl(G^{-1} E\bigr)^k \\[8pt]
&\=\exp\bigl(\bar ee\,\pa_{\bar gg}\bigr)\,\tr\,\Bigl\{
\ln G\,-\,\bar ee\,\nwarrow\swarrow\,-\ (\bar ee)^2\,\bigl(\sfrac12
 \nwarrow\swarrow\nwarrow\swarrow
+\nwarrow\nwarrow\swarrow\swarrow
\bigr)\,-\,(\bar ee)^3\,\times \\[4pt]
&\quad\times\,\bigl(\sfrac13
 \nwarrow\swarrow\nwarrow\swarrow\nwarrow\swarrow
+\nwarrow\nwarrow\nwarrow\swarrow\swarrow\swarrow
+\nwarrow\nwarrow\swarrow\nwarrow\swarrow\swarrow
+\nwarrow\swarrow\nwarrow\nwarrow\swarrow\swarrow
\bigr)\,+\ldots\Bigr\}\ .
\eal
\ee
The operator $\ln(\unity+G^{-1}E)$ is merely Dixmier class while $G^{-1}E$ is of order~$2$. 
The logarithmic divergence of its trace can be calculated as the Wodzicki residue of $G^{-1}E$ which is~$0$. 
Thus the first term of the sum in~\eqref{ringpotential} vanishes and its sum evaluates to the trace 
of $\ln(\unity+G^{-1}E)-G^{-1}E$, which is of order~$4$ and hence trace class.
The trace produces calculable infinite sums of rational functions of~$n$. 
Since the summands of order $(\bar ee)^k$ decay as $n^{-2k}$ for large~$n$,
all sums converge, except for the leading $\tr\ln G$ term.
A remedy consists in subtracting an infinite constant,\footnote{
In zeta-function regularization this constant equals to $\ln\pi$.}
\bea \non
\sum_{n=0}^\infty\ln\bigl(\bar gg+(n{+}2)(n{+}1)\bigr) & \quad\longrightarrow\quad &
\sum_{n=0}^\infty\Bigl[\ln\bigl(\bar gg+(n{+}2)(n{+}1)\bigr)\ -\ \ln\bigl((n{+}2)(n{+}1)\bigr)\Bigr]\\
& \quad = & \sum_{n=0}^\infty \ln\Bigl[1+\frac{\bar gg}{(n{+}2)(n{+}1)}\Bigr] \label{leadingterm} \\
& \quad = & \ln\cos W \ -\ \ln\bar gg \ -\ \ln\pi \qquad\textrm{with}\qquad
W \= \sfrac\pi2\sqrt{1{-}4\bar gg}\ . \non
\eea
We display the first few terms of this power series expansion in $\bar e e$:
\be \label{ncring}
\bal 
\frac{\cal K}{16\pi\th} \ \ &\!\!=\ \;
\ln\cos W \ -\ \ln\bar gg \ -\ \ln\pi \ +\ 
(\bar ee)^1\,\pi^2\sfrac{\bar gg}{4\bar gg{+}3}\,\sfrac{\tan W}{W} \\[8pt]
&\!+\ (\bar ee)^2\pi^4\Bigl\{
\sfrac{48(\bar gg)^4{+}200(\bar gg)^3{-}33(\bar gg)^2{+}27\bar gg}
{4\,(4\bar gg{+}3)^3(4\bar gg{+}15)}\,\sfrac{\tan W}{W^3}\ -\
\sfrac{(\bar gg)^2}{2\,(4\bar gg{+}3)^2}\,\sfrac{\sec^2W}{W^2}\Bigr\} \\[8pt]
&\!+\ (\bar ee)^3\pi^6\Bigl\{
\sfrac{10240(\bar gg)^8+171520(\bar gg)^7+878336(\bar gg)^6+\ldots
-13770(\bar gg)^2+6075\bar gg}
{8\,(4\bar gg{+}3)^5(4\bar gg{+}15)^2(4\bar gg{+}35)}\,
\sfrac{\tan W}{W^5} \\[4pt] 
&\qquad\qquad -
\sfrac{48(\bar gg)^5+200(\bar gg)^4-33(\bar gg)^3+27(\bar gg)^2}
{4\,(4\bar gg{+}3)^4(4\bar gg{+}15)}\,\sfrac{\sec^2W}{W^4}\ +\
\sfrac{(\bar gg)^3}{3\,(4\bar gg{+}3)^3}\,\sfrac{\tan W\sec^2W}{W^3}
\Bigr\} \ +\ \ldots \ .
\eal
\ee
The $g$~dependence is exact in each order in~$\bar ee$. 
Note that all powers of~$\frac{\bar ee}{\bar gg}$ from the expansion of $-\ln(\bar gg+\bar ee)$
get cancelled.  For $\bar gg>\frac14$ one must
analytically continue~$W$, and the trigonometric functions convert to hyperbolic ones.
When $g\to0$, we find
\be
\ln\cos W\ -\ \ln\bar gg \= \ln\pi\ +\ \bar gg\ +\ (\sfrac32{-}\sfrac{\pi^2}{6})(\bar gg)^2\ +\ \ldots
\und \bar gg\sfrac{\tan W}{W}\= \sfrac{2}{\pi^2}\ +\ \ldots\ ,
\ee
and so the whole expression is regular at $e=g=0$ and behaves as \ $\bar gg+\frac23\bar ee+\ldots$. 
Keeping $|\sfrac e g|\ll1$ fixed, we can vary~$\th$: The $\th\to\infty$ limit ($g,e\to0$)
is smooth since $\bar gg\tan W\sim\bar g g\sec W$ remain finite; for weak noncommutativity
($g,e\to\infty$) one indeed recovers the ring-regime expansion of the commutative 
result~(\ref{Kringcom}), \ ${\cal K}\={\cal K}_0 + O\bigl(\sfrac{\th^2}{|\g|}\bigr)$.

\section{Deformed lumps -- the perturbation approach}

\noindent
For $|g|\ll|e|$ we encounter the two-lump regime. We should like to set up a perturbation expansion
in powers of~$|g|^2$ here. Formally, it reads
\be
\frac{\cal K}{16\pi\th}\=
\tr\,\ln\bigl[ F(e)+\bar gg \bigr] \=
\tr\,\ln F(e)\ +\ \tr\,\ln\bigl[ \unity+\bar gg\,F(e)^{-1}\bigr]\ .
\ee
It is easy to see that the operator $F(e)$ has no zero modes.
Because it is an elliptic operator in the Shubin calculus, 
its spectrum is non-degenerate and discrete.
While the order of~$F(e)$ is~$4$, the operator $\ln\bigl[\unity+\bar gg\,F(e)^{-1}\bigr]$ is trace class. 
The eigenvalues of $F(0)$ obviously are $\la_n(0)=(n{+}2)(n{+}1)$ for $n=0,1,2,\ldots$.
The eigenvalues $\la_n(e)$ for are called `spheroidal'~\cite{lema,abramowitzstegun}.
Hence, for small $\bar gg$, we can write a formal series expansion
where each term is exact in~$e$:
\be \label{lumpexp}
\frac{\cal K}{16\pi\th} 
\= \sum_{n=0}^\infty \ln \bigl( \bar gg+\la_n(e) \bigr) 
\= - \sum_{k=1}^\infty \sfrac1k (-\bar gg)^k\sum_{n=0}^\infty\la_n(e)^{-k}\ . 
\ee

The spheroidal eigenvalues admit a Taylor expansion in $|e|^2$,
\bea \non
\la_n(e) &=& (n{+}2)(n{+}1)\,\Bigl\{ 1\,+\,\frac{2}{(2n{+}1)(2n{+}5)}\bar ee\,+\,
\frac{2\,(4n^4+24n^3+13n^2-69n+1)}{(2n{-}1)(2n{+}1)^3(2n{+}5)^3(2n{+}7)}
(\bar ee)^2\,+\,\ldots \Bigr\} \\[4pt]
&=:& (n{+}2)(n{+}1)\ \bigl\{ 1\ +\ \hat{\la}_n(e)\bigr\}\ . \label{spheroidal}
\eea
Using the above factorization, we can rearrange (\ref{lumpexp}) to obtain
\be \label{nclump}
\bal
\frac{\mathcal{K}}{16\pi\theta}&\=
\sum_{n=0}^\infty \ln\bigl[\bar gg+(n{+}2)(n{+}1)\bigr]\ +\
\sum_{n=0}^\infty \ln\bigl[1+\sfrac{(n{+}2)(n{+}1)}{\bar gg+(n{+}2)(n{+}1)} \hat{\la}_n(e)\bigr] \\[8pt]
&\=\textrm{const}\ +\ \ln\cos W\ -\ \ln\bar gg\ -\
\sum_{k=1}^\infty\sfrac{(-1)^k}{k} \sum_{n=0}^\infty
\bigl(1+\sfrac{\bar gg}{(n{+}2)(n{+}1)}\bigr)^{-k} {\hat{\la}_n(e)}^k\ .
\eal
\ee

Expanding in powers of~$|e|^2$  and performing the $n$-sums, 
we get agreement with (\ref{ncring}) expanded in powers of~$|g|^2$.
For the strong noncommutative limit $\th\to\infty$, we can thus provide
a double Taylor expansion:
\be
\label{perturbedlump}
\frac{\mathcal{K}}{16\pi\theta}\= 
\sfrac23 |e|^2\ +\ |g|^2\ -\ \sfrac{4}{45} |e|^4\ -\ 
\sfrac29 |e|^2|g|^2\ +\ (\sfrac32{-}\sfrac{\pi^2}{6}) |g|^4\ +\ \ldots\ .
\ee
An expansion analogous to (\ref{ncring}), valid also for large~$|e|^2$ in the two-lump regime,
requires an analytic formula for~$\la_n(e)$. The best we can hope for is an asymptotic expansion
around $e=\infty$ by means of semiclassical techniques. Such a result would also allow us to connect
with the commutative limit $\th\to0$.

\newpage

\section{Deformed lumps -- the zeta function approach}

\noindent
In this section we focus on calculating the potential for the noncommutative Ward metric 
in the two-lump region $|\g|\ll|\eps|$ and especially for the strongly noncommutative limit $\th\to\infty$. 
We will do this by finding the $\zeta$-function of the fourth-order operator $T^\dagger T$. 

To calculate the zeta function of $T^\dagger T$ we first evaluate its heat trace. We have that
\be
\bal
\tr(\mathrm{e}^{-tT^\dagger T})&\= \sum_{n=0}^\infty\<n|\mathrm{e}^{-t\,T^\+T}|n\> \=\sum_{n=0}^\infty\langle n|\mathrm{e}^{-t\left((a^2+z)(a^{\dagger 2}+w)+|g|^2\right)}|n\rangle\big|_{z=\bar{w}=e} \= \\
&\=\sum_{n=0}^\infty\sum_{k=0}^{\lfloor \frac{n}{2}+1\rfloor} \sfrac{|e|^{2k}(-n-2)_{2k}t^{2k}}{(k!)^2}\,
\mathrm{e}^{-t\left((n+2)(n+1)+|e|^2+|g|^2\right)}\ ,
\eal
\ee
where we in the last equality use the Taylor expansion of the holomorphic function 
$(z,w)\mapsto \sum _{n=0}^\infty \langle n|\mathrm{e}^{-t\left((a^2+z)(a^{\dagger 2}+w)+|g|^2\right)}|n\rangle$ and employ
the Pochhammer symbol $(x)_k=x(x{+}1)\cdots (x{+}k{-}1)$ if $k>0$ and $(x)_0=1$. With the help of
\be
\int_0^\infty\!\diff t\ t^{s+k-1}\mathrm{e}^{-\sigma t}\=\sigma^{-s-k}\Gamma(s+k) \und
\Gamma(s+k)\=(s)_k\,\Gamma(s)
\ee
and splitting off the $k{=}0$ terms,
it follows that for $\Re(s)>\sfrac12$ the zeta function of $T^\dagger T$ is given by
\be \label{thezeta}
\zeta_{T^\dagger T}(s)\=\sum_{n=0}^\infty \bigl((n{+}2)(n{+}1)+|e|^2+|g|^2\bigr)^{-s}+
\sum_{n=2}^\infty\sum_{k=1}^{\lfloor \frac{n}{2}\rfloor} \sfrac{|e|^{2k}(-n)_{2k}(s)_{2k}}{(k!)^2}
\bigl(n(n{-}1)+|e|^2+|g|^2\bigr)^{-s-2k}\ .
\ee
This expression is valid for all values of $e$ and $g$, but it is difficult to treat for most of the values. 
Applying the functional equation~(\ref{functionalrelation}) for $|g|^2$ and integrating ditto again 
from $0$ to $|g|^2$ we arrive at the expression
\be \label{thedeterminant}
\bal
\frac{\mathcal{K}}{16\pi\theta}&\=-\zeta_{T^\dagger T}'(0)|_{g=0}\ +\ 
\sum_{n=0}^\infty \ln\Bigl[1+\sfrac{|g|^2}{(n{+}2)(n{+}1)+|e|^2}\Bigr] \\
&\ -\ \sum_{n=2}^\infty\sum_{k=1}^{\lfloor \frac{n}{2}\rfloor} \sfrac{|e|^{2k}(-n)_{2k}(2k)!}{(k!)^2}
\Bigl[\bigl(n(n{-}1)+|e|^2+|g|^2\bigr)^{-2k}-\bigl(n(n{-}1)+|e|^2\bigr)^{-2k}\Bigr] \ .
\eal
\ee
This representation of the potential $\mathcal{K}$ is valid for all values of the moduli parameters. 
Observe that the third term vanishes at $g=0$ as well as at $e=0$. 
We shall calculate the first and the second term explicitly. 
The first term can be calculated using a result of Lesch \cite{lesch} that is a variation of the celebrated 
Gelfand-Yaglom theorem. The second term can be evaluated using~\eqref{leadingterm}. 
Finally we show that the remainder, i.e.\ the third term, vanishes to fourth order at the origin 
when fixing $\frac{|g|}{|e|}$. This will produce an asymptotic expansion up to third order 
for the strongly noncommutative limit $\th\to\infty$ in the two-lump case.

\subsection{The first term} 

\noindent
Let us turn towards the calculation of $-\zeta_{T^\dagger T}'(0)|_{g=0}$. 
Recall the notation $F(e)=T^\dagger T|_{g=0}$. 
We observe that the heat trace of~$F(e)$ is independent of our choice to consider the antisoliton operator 
$(a^2{+}e)(a^{\+2}{+}\bar{e})$ instead of the soliton operator $(a^{\+2}{+}\bar{e})(a^2{+}e)$, 
while their nonzero spectra coincide.  
It was was proven in~\cite{lema} that the operator $F(e)$ for the \emph{soliton choice} is, 
via multiplication by $\mathrm{e}^{x^2/2}$, Fourier transformation and a change of variables, 
equivalent to the singular Sturm-Liouville operator
\be
\tilde{F}(e)\ :=\ -\partial_z(1-z^2)\partial_z\ +\ \frac{1}{1-z^2}\ +\ |e|^2(1-z^2)
\qquad\textrm{for}\quad z\in[-1,1]\ .
\ee
For details, see \cite{lema}. The operator $\tilde{F}(e)$ is well studied and has discrete spectrum. 
Its eigenvectors are known as the oblate spheroidal wave functions~\cite{abramowitzstegun} 
with the eigenvalues \eqref{spheroidal}. 
We improve the problem further by another change of variables, $z=\sin(x)$, 
which transforms the eigenvalue problem to solving
\be
-\partial_x^2h(x)\ +\ \bigl(\mathrm{cosec}(x)+|e|^2\cos(x)\sin^2(x)-1\bigr)h(x)\=\lambda\,h(x)
\qquad\textrm{for}\quad x\in[-\sfrac\pi2,\sfrac\pi2]\ ,
\ee
which is a regular-singular Sturm-Liouville problem. 
The solution normalized at $\pm\sfrac\pi2$ is given by
\be
\phi_\pm(x)\ :=\ |e|^{-1}\sinh(|e|(\sin(x)\mp 1))\ .
\ee
In particular, the Wronskian of this equation is
\be
{\cal W}(\phi_-,\phi_+)\ \equiv\ \phi_+'\phi_--\phi_+\phi_-'\= |e|^{-1}\sinh(2|e|)\ .
\ee
The determinants of regular-singular Sturm-Liouville problems have been studied in~\cite{lesch}, 
and it follows that
\be
\det\bigl[F(e)\bigr]\=2|e|^{-1}\sinh(2|e|)
\qquad\Longrightarrow\qquad
-\zeta_{T^\dagger T}'(0)|_{g=0}\= \ln \frac{2\sinh(2|e|)}{|e|}\ ,
\ee
which is consistent with the calculation in~\cite{lema} based on the Gelfand-Yaglom theorem.

\subsection{The second term}

\noindent
This term derives from the first term of~\eqref{thezeta}. Let us denote
\be
\zeta^0(s,t) \= \sum_{n=0}^\infty \bigl((n{+}2)(n{+}1)+t\bigr)^{-s} \ ,
\ee
so if $t\geq t'>0$ it follows from integrating the functional equation \eqref{functionalequation} 
from $t'$ to $t$ that 
\be
\sum_{n=0}^\infty \ln \Bigl[ 1 + \frac{t-t'}{(n{+}2)(n{+}1)+t'}\Bigr] \=
\sum_{n=0}^\infty \ln\frac{(n{+}2)(n{+}1)+t}{(n{+}2)(n{+}1)+t'} \=
-\pa_s \zeta^0(s,t)|_{s=0}\ +\ \pa_s \zeta^0(s,t')|_{s=0} \ .
\ee
Comparing with the result~(\ref{leadingterm}),
\be
-\pa_s \zeta^0(s,t)|_{s=0} \= \sum_{n=0}^\infty \ln\bigl((n{+}2)(n{+}1)+t\bigr)
\= \ln\cos W(t)\ -\ \ln t\ +\ \textrm{const}\ ,
\ee
and putting $t=|e|^2+|g|^2$ and $t'=|e|^2$, we conclude with $W(t)=\sfrac\pi2\sqrt{1-4t}$ that
\be \label{seconddisc}
\sum_{n=0}^\infty \ln\Bigl[1+\frac{|g|^2}{(n{+}2)(n{+}1)+|e|^2}\Bigr] \=
\ln \frac{|e|^2\,\cos\bigl(\sfrac\pi2\sqrt{1-4|e|^2-4|g|^2}\bigr)}
{\bigl(|e|^2+|g|^2\bigr)\,\cos\bigl(\sfrac\pi2\sqrt{1-4|e|^2}\bigr)}
\ee
in the disk $|g|^2+|e|^2<\sfrac14$. This result is finite at $e=0$ and
extends to the outside of the disk by analytic continuation, 
as was already mentioned.

\subsection{The remainder term and the $\theta\to \infty$ asymptotics}

\noindent
Let us finally consider the remainder term in~(\ref{thedeterminant}), which is given by 
\be
\bal
R(e,g)\ &:=\ 
\sum_{n=2}^\infty\sum_{k=1}^{\lfloor \frac{n}{2}\rfloor} \sfrac{|e|^{2k}(-n)_{2k}(2k)!}{(k!)^2}
\Bigl[\bigl(n(n{-}1)+|e|^2\bigr)^{-2k}-\bigl(n(n{-}1)+|e|^2+|g|^2\bigr)^{-2k}\Bigr] \\
&\= \sum_{j=1}^\infty |e|^{2j}\,R_j\bigl(|\sfrac{g}{e}|^2\bigr)\ .
\eal
\ee
One can calculate $R_j$ by means of a Taylor expansion of $R$. 
This is relevant for $\th\to\infty$ because fixing the dimensionful moduli 
implies that $e$ and $g$ are $O(\th^{-1})$. One obtains the series
\be
R_j(v)\=\sum_{n=2}^\infty\sum_{k=1}^{\min\left(j,\left\lfloor \frac{n}{2}\right\rfloor\right)} 
\frac{2k(-n)_{2k}(k+j-1)!}{k!}\ \frac{(-1)^{k-j}\bigl(1-(1+v)^{j-k}\bigr)}{n^{2k}\,(n{-}1)^{2k}}\ .
\ee
It is a polynomial of degree $j{-}1$ vanishing at $v=0$. The first two of the $R_j$s are given by 
\be
R_1(v)\=0 \und R_2(v)\=12v\ .
\ee

We conclude from \eqref{thedeterminant} that in the two-lump case the complete K\"ahler potential reads
\be
\bal
\frac{\mathcal{K}}{16\pi\theta}&\= \ln\frac{2\sinh(2|e|)}{|e|}\ +\
\ln \frac{|e|^2\,\cos\bigl(\sfrac\pi2\sqrt{1{-}4|e|^2{-}4|g|^2}\bigr)}
{\bigl(|e|^2{+}|g|^2\bigr)\,\cos\bigl(\sfrac\pi2\sqrt{1{-}4|e|^2}\bigr)}\ +\ 
\sum_{j=2}^\infty |e|^{2j} R_j\bigl(|\sfrac{g}{e}|^2\bigr) \\[8pt]
&\= \textrm{const}\ +\
\bigl(\sfrac23+|\sfrac{g}{e}|^2\bigr)|e|^2\ +\ O\bigl(|e|^4\bigr)\ ,
\eal
\ee
where the second line shows its $e\to 0$ asymptotics.
So in the limit $\theta\to \infty$ the asymptotics is
\be
\mathcal{K}\=
\frac{4\pi}{\th}\bigl(\sfrac{2}{3}|\eps|^2+|\g|^2\bigr)\ +\ O(\th^{-3})\ ,
\ee
in agreement with the naive double Taylor expansion~(\ref{perturbedlump})
of the expression~(\ref{ncring}) in the ring regime.

\section{Conclusions}

\noindent
We have investigated the structure of the noncommutative deformation of the Ward metric in the charge-$2$ sector, as a function of the two dimensionless complex moduli~$e=\frac{\eps}{2\th}$ and $g=\frac{\g}{2\th}$. Along the curve $e=0$ in the moduli space the deformed potential was expressed in closed form. The same was done using a Gelfand-Yaglom-type result along the limiting curve $g=0$, which is not inside the classical moduli space. The $\zeta$-function and the noncommutative deformation of the Ward metric in the charge-$2$ sector was expressed as an explicit power series.

In the ring-like regime $|e|\ll|g|$, the potential is controlled by perturbation theory and admits the calculation of asymptotics. Here, the classical Ward potential is recovered in the commutative limit, at least to order~$|\frac{e}{g}|^8$. 

The two-lump regime $|e|\gg|g|$ provides a bigger challenge. There we employed both perturbation theory and a more direct approach using the $\zeta$-function to obtain asymptotics for the strong noncommutative limit. The results interpolates between the values along the curves $e=0$ and $g=0$. So it appears that the two methods lead to coinciding results.

Unfortunately, direct approaches like $\zeta$-functions and perturbation theory does not seem to reach the commutative limit of the deformed lumps. As of present, we know only the form of the leading noncommutative correction~\cite{lema},
\be
\frac{\cal K}{16\pi}\= |\eps|\,A_0\bigl(|\sfrac{\g}{\eps}|^2\bigr)
\ +\ \th\,\mathrm{e}^{-2|\eps|/\th} A_1\bigl(|\sfrac{\g}{\eps}|^2\bigr) 
\ +\ O\bigl(\th\,\mathrm{e}^{-4|\eps|/\th},\sfrac{\th^2}{|\eps|}\bigr)\ ,
\ee
where $16\pi|\eps|A_0={\cal K}_0$, and $A_1$ is an unknown function.
There exist some very interesting semiclassical techniques for the parameters approaching infinity~\cite{burgheleafriedlanderkappeler}. With their help one may write down a full asymptotic expansion of $\mathcal{K}$ in the commutative limit in terms of local invariants given by integrals of certain rational functions coming from the Shubin calculus. Even though this involves elliptic integrals, we have not reached an exact expression and leave this problem for future work.

\newpage

\begin{figure}
\includegraphics[width=15cm]{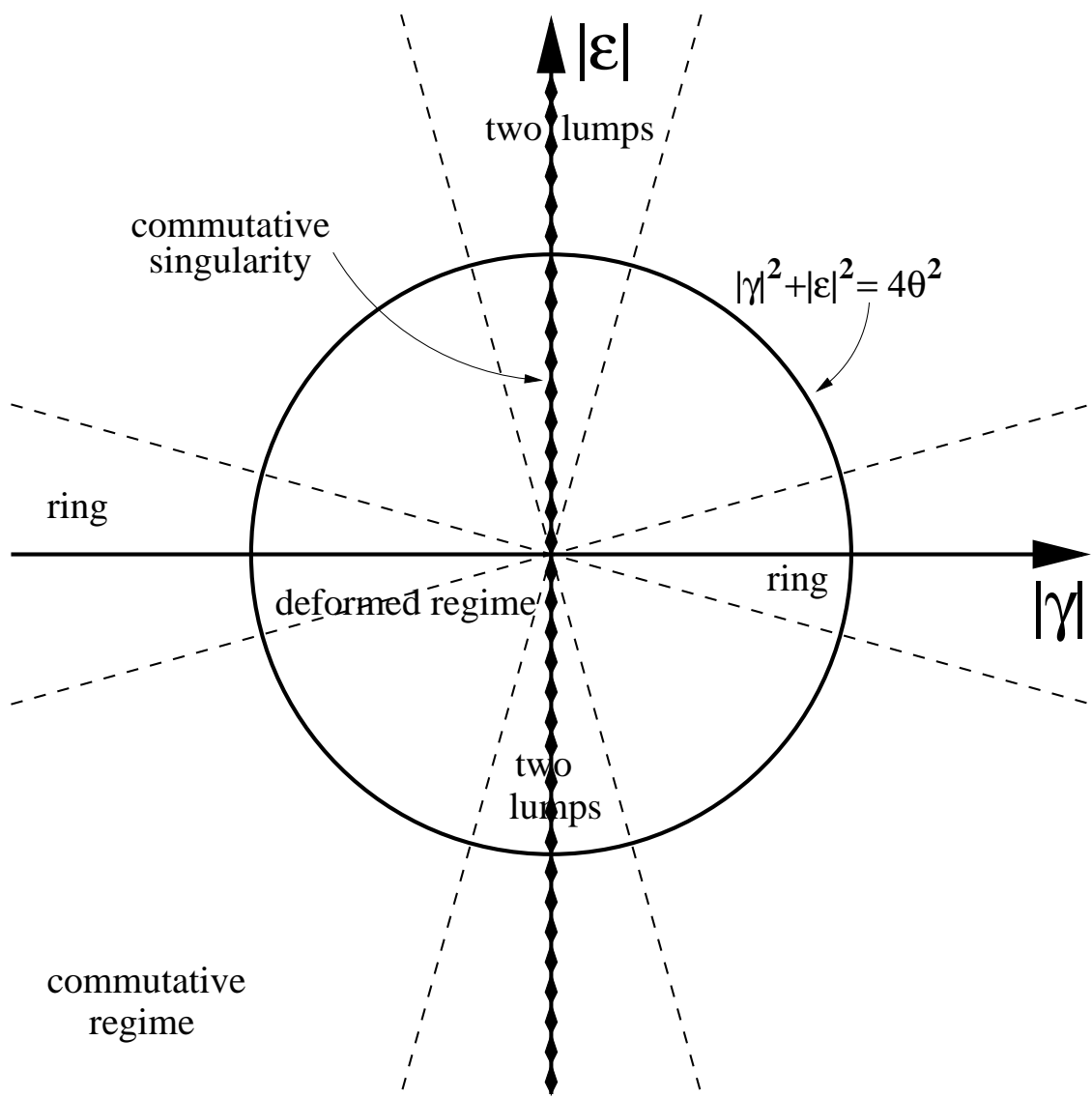} 
\caption{modulus-of-moduli space}
\label{fig2}
\end{figure}

\end{document}